\documentstyle[prb,aps,multicol,epsfig]{revtex}

\begin{document}

\title{Non-equilibrium dynamics in a 3d spin-glass}

\author{T. Jonsson, K. Jonason, P. J\"{o}¥nsson, and P. Nordblad}

\address{
Department of Materials Science, Uppsala University, Box 534, SE-751 21
Uppsala, Sweden}

\maketitle

\begin{abstract}
Non-equilibrium dynamics in a Ag(11\%Mn) spin-glass has been studied by low 
frequency ac-susceptibility and magnetic relaxation experiments. The results
unequivocally show that spin structures that memorize the cooling process are 
imprinted in the system. These imprinted structures disclose themselves through 
dramatic changes of the dynamics on re-heating the spin-glass through the 
temperatures where intermittent stops or changes of the cooling rate have been 
imposed. We can qualitatively  interpret our results in terms of the droplet 
spin-glass model developed by Fisher and Huse [Phys. Rev. B {\bf 38} (1988) 
373; 386].
\end{abstract}
\begin{multicols}{2}
\narrowtext

\section{Introduction}
A  successful real space theory for the dynamics of 3d spin-glass systems is 
the droplet model developed a decade ago by Fisher and Huse. 
\cite{FisherHuse} Particular 
experimental characteristics such as the aging phenomenon \cite{aging} and the 
non-existence of a phase transition in a magnetic field  \cite{field} are 
inherent 
properties of the droplet spin-glass model. However, the validity of this 
model has been questioned: e.g. a compact domain growth picture is not 
consistent with the multiple memory effect recently found in 3d spin-glass 
systems. \cite{Nordblad,Jonason} Fig. 1 shows a low frequency 
ac-susceptibility experiment on 
a Cu(Mn) spin-glass where this suggestive effect is illustrated. In this 
paper, extended measurements of some associated non-equilibrium features of 
the ac-susceptibility and the dc-magnetic relaxation  on the 3d spin-glass 
material Ag(11at\% Mn) are presented. The results can qualitatively be 
incorporated in the original droplet model \cite{FisherHuse} without apparent 
contradictions 
and we do not here try to interpret our results on aging in spin-glasses in 
terms of non real space models. \cite{Bouchaud}

\section{The droplet model}
The results of this study are interpreted in the spirit of the droplet model 
using relevant parts as summarized in the following points;

\begin{itemize}

\item[i)] The spin-glass ground state is unique but two-fold degenerate by 
its spin reversal symmetric state.  

\item[ii)] Chaos with temperature - a small temperature shift changes the ground 
state spin configuration completely on long enough length scales. 
\cite{BrayMoore}
\item[iii)] The length scale, $l$, up to which no essential change in spin 
configuration is observed after a temperature change $\Delta T_{*}¥$ 
introduces the concept of an overlap length $l(\Delta T_{*}¥)$.
\item[iv)] In the non-equilibrium case, the development towards the ground state 
is governed by the growth of domains belonging to either of the two 
degenerate ground states. The typical domain size after a time, $t_{w}$, at a 
constant temperature $T$, is 

\begin{equation}
	R \propto \left({\frac{T \ln(t_{w}¥ / \tau_{0}¥ ) }{\Delta (T) } } 
	\right)^{1 / \psi}  
\label{eq:pa1}
\end{equation}	
where $\tau_{0}¥$ is a microscopic time of order $10^{-13}$ s. $\Delta 
(T)$ sets the free energy scale and $\psi$ is a barrier exponent. This 
length scale R, that we often will refer to as a typical domain size, 
should not be taken literally, but rather be considered as a typical 
measure of the smallest separation between domain walls remaining after 
the wait time $t_{w}¥$.
\end{itemize}

In order to discuss the relaxation towards equilibrium, a 3d Ising 
spin-glass quenched to a temperature $T_{1}¥ < T_{g}¥$ is considered. The 
state at 
$t = 0$ is hence characterized by spins of random direction. To further 
simplify the following argumentation it is also assumed that the spins 
are situated on a regular lattice. In the droplet model 
\begin{figure}
	\centerline{\epsfig{width=8.5truecm,angle=0,file=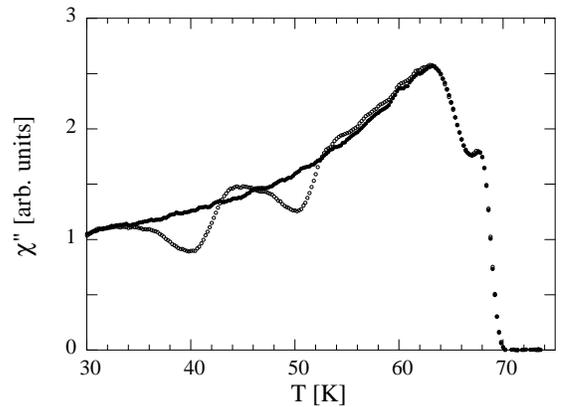}}
	\caption{The out-of-phase component of the ac susceptibility of a
	Cu(13.5 at\% Mn) ($T_{g}¥ \approx 68$ K) spin-glass measured on continuous 
	heating after the sample has been subjected to two intermittent 
	stops of order $10^{4}$ s at 40 and 50 K during the cooling procedure. 
	The two pronounced dips in the curve occur at the temperatures of 
	the stops at constant temperature during cooling. $f = 1.7$ Hz. }
\end{figure}

\noindent there exists 
only two degenerate ground states, $\it{\Psi}$ and its spin reversal counterpart
$\it{\tilde{\Psi}¥}$, 
it is thus possible to map all spins to either of the two desired ground 
states immediately after a quench. Citing Fisher and Huse 
\cite{FisherHuse} ``This results 
in an interpenetrating network of regions of the two states''. In fact, the 
domain walls of one lattice spacing, which separates the two states 
$\it{\Psi}$ and $\it{\tilde{\Psi}¥}$   
at $t=0$ after the quench should have a fractal surface giving rise to fractal 
domains of all sizes. The subsequent equilibration process at $t > 0$ is 
governed by droplet excitations yielding a domain growth to a typical length 
scale which depends on temperature and wait time, $t_{w1}¥$, according to 
Eq. \ref{eq:pa1}. 
After this time, fractal structures typically smaller than $R(T_{1}¥, 
t_{w1}¥)$ have 
become equilibrated, but, viewing the system on a larger length scale, it 
still looks intact. The implication of this argument is that large enough 
fractal structures which were obtained at $t=0$ after the quench are virtually 
unaffected after the wait time $t_{w1}¥$. If the system thereafter is 
quenched to a 
considerably lower temperature $T_{2}¥$, a new fractal domain structure can be 
mapped onto the equilibrium configuration at $T_{2}¥$. The short length 
scale limit 
is now set by the overlap length, $l(T_{1}¥ - T_{2}¥)$, and fractal 
structures will be 
Ôwashed outÕ with increasing wait time, $t_{w2}¥$, starting from this length 
scale 
and ending at the size $R(T_{2}¥, t_{w2}¥)$. Structures on longer length scales 
are 
again intact. If the system is further quenched and then instantly re-heated 
to $T_{2}¥$, the domain structure at $T_{2}¥$ of course remains 
unaffected, but also, 
when instantaneously heated back to $T_{1}¥$, the re-structuring on short length 
scales, $R(T_{2}¥,t_{w2}¥) \ll R(T_{1}¥,t_{w1}¥)$, that occurred at 
$T_{2}¥$ is rapidly washed out and 
the system is effectively left with only the original large length scale 
domain structure which was imprinted at the first quench to $T_{1}¥$. If 
the fractal 
droplet model is relevant, these basic properties of the logarithmically 
slow (eternal) equilibration process should be reflected in the dynamic 
response of a real 3d spin-glass subjected to a corresponding thermal history. 
In a real experimental situation, the system can only be cooled and heated 
at a controlled but finite rate. The domain growth is in such processes 
governed by the cooling/heating rate and limited by an interplay between 
the domain growth (Eq. \ref{eq:pa1}), the chaotic nature of the spin 
glass phase and 
the overlap between states on short enough length scales. Accounting for 
this complication, we have designed a series of dynamic magnetization and 
susceptibility experiments that should mirror the discussed development of a 
domain structure in an equilibrating 3d spin-glass. 

Droplet excitations  occur on time limited length scales and are in fact 
responsible for the equilibration process of the quenched spin-glass, also 
giving rise to the domain growth function Eq. \ref{eq:pa1}. If a weak 
magnetic field  
is applied on the spin-glass at time $t_{obs}¥ = 0$, the time 
dependent response, $m(t_{obs}¥)$, is due to a continuous magnetization 
process governed by polarization of droplets of size, $L(t_{obs}¥)$: 

\begin{equation}
	{L} \propto \left({\frac{T \ln(t_{obs}¥ / \tau_{0}¥ ) }{\Delta (T) } } 
	\right)^{1 / \psi}
\label{eq:pa2}
\end{equation}

Since $L$ grows with the same logarithmic rate as $R$, the relevant droplet 
excitations and the actual domain size become comparably large at time 
scales $t_{w}¥ \approx t_{obs}¥$. For $t_{obs}¥ \ll t_{w}¥$ 
the relevant excitations occur mainly 
within equilibrated regions while for $t_{obs}¥ \gg t_{w}¥$ these excitations 
occur on length scales of the order of the growing domain size and 
yield a non-equilibrium response. A crossover occurs in the 
intermediate region, $t_{w}¥ \approx t_{obs}¥$ which is characterized 
by a maximum in the
relaxation rate 
${S(t)}={1/H \; \partial m(t) / \partial \ln(t)}$. 
This aging behavior is also reflected in low frequency 
ac-susceptibility measurements. Different from dc-relaxation measurements 
is that the observation time is kept constant by the probing frequency, 
$\omega / {2  \pi}$, according to $t_{obs}¥=1 / \omega$ and hence, 
in an experiment at constant 
temperature, sets the associated probed length scale, $L(T ,1/\omega)$, fixed. 
In this study we are primarily interested in the processes that 
affects the response at the observation time of the experiment, 
the consequences of these are best exposed in plots of the relaxation 
rate $S$ vs. $t_{obs}¥$ for different wait times, the out-of-phase component of 
the ac-susceptibility  vs. time at constant temperature or vs. 
temperature at different cooling/heating rates.  Therefore, the dc-relaxation 
measurements are primarily presented in the form of the relaxation 
rate, $S$, 
and the ac-susceptibility is represented by its imaginary part, $\chi''$, these 
quantities  can be related through the relation, \cite{JMMM}

\begin{equation}
	\chi'' \approx \frac{\pi}{2} S \left( t_{obs}¥ \right), \; \; \; \;
	t_{obs}¥ = 1 / \omega
	\label{eq:pa3}
\end{equation}	

\section{Experimental}
The sample was prepared by melting pure Ag and Mn together at 
$T$ = 1000 ${}^\circ$C in an 
evacuated atmosphere. After annealing the sample at 850 ${}^\circ$C for 72h it was 
water quenched to room temperature.

The experiments have been performed in a non-commercial low-field SQUID 
magnetometer. \cite{Magnusson} The dc magnetic field is generated by a small 
superconducting solenoid always working in persistent mode during 
measurements. The ac-field was generated by a copper coil directly wound 
on the sample which is shaped as a 5 mm long cylinder, 2.5 mm in 
diameter. The magnetic response of the sample, subtracted by the 
ac-field from a compensating coil, was recorded with a set of pick-up 
loops positioned to form a third order gradiometer. At the position of 
the sample, the resulting rms value of the ac-field was 0.1 Oe and 
the background field was less than 1 mOe. All ac-susceptibility 
measurements were performed at a frequency of 1.7 Hz. A sapphire 
rod was used to provide a good thermal contact between the heater, 
the thermometer and the sample.

\section{Results and discussion}

\subsection{Aging characteristics}
Figs. 2 and 3 introduce the overall behavior of the dynamic susceptibility 
of our Ag(11at\%Mn) sample and expose two classical manifestations of the 
aging phenomenon in spin-glasses. In the inset to Fig. 2 both components 
of the ac-susceptibility, $\chi (\omega) = \chi' (\omega) + \chi'' 
(\omega)$ at the frequency $\omega / 2 \pi$ = 1.7 Hz, 
are plotted vs. temperature at a cooling rate of 0.25 K/min. 
The maximum in $\chi' (T)$, defining the freezing temperature, is accompanied 
by the onset of an out-of phase component of the susceptibility. 
In the main frame of Fig. 2 the influence of aging on $\chi''$ is illustrated 
by a plot of $\chi''$ vs. time elapsed at 23 K, where the continuous cooling 
was interrupted. Fig. 3 exemplifies the results from an ordinary zero 
field cooled (ZFC) aging experiment at $T_{m}¥ = 27$ K, in (a) the 
magnetization  
and in (b) the corresponding relaxation rate is plotted vs. the 
observation time. The sample is probed in a dc-field  of 1 Oe after the 
wait times $t_{w}¥$ = 100 s (open symbols) and $t_{w}¥$ = 3 000 s 
(filled symbols) at $T_{m}¥$.  
The signatures of aging in spin-glasses: an inflection point in $m(t)$  
and a corresponding maximum in $S(t)$ vs. $\log t_{obs}¥$ are reproduced 
in the figure.

Some words about time and non-equilibrium dynamics: experimentally we are 
confined to time scales set by the cooling/heating rates and the time it 
takes to thermally equilibrate the sample to a constant temperature. 
Although spin-glass dynamics occurs on all time scales ranging from 
$10^{-13}$ s to infinity, the shortest observation times where consequences 
of non-equilibrium dynamics can be observed experimentally  are confined 
to $t_{obs}¥ \geq 10^{-3} t_{aeff}¥$ where $t_{aeff}¥$ is an effective 
age of the system set 
by the cooling/heating rate or the time spent 

\begin{figure}
	\centerline{\epsfig{width=8.5truecm,angle=0,file=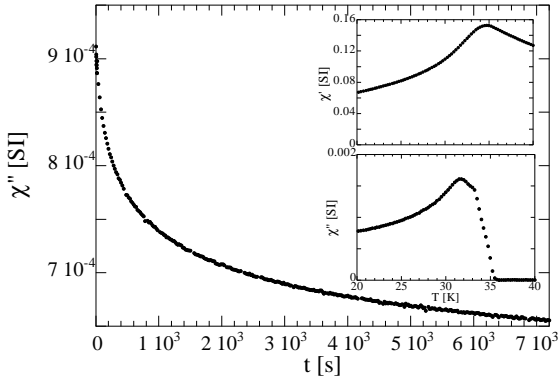}}
	\caption{Relaxation of $\chi'' (t)$ with time at 
	constant temperature after the sample has been cooled from a
	temperature above $T_{g}¥$ to $T = 23$ K.
	Inset, both components of the ac-susceptibility vs. $T$ measured at 
	a cooling rate of 0.25 K/min. $f = 1.7$ Hz, $h_{ac}¥ = 0.1$ Oe,
	Ag(11 at\% Mn).}
\end{figure}

\noindent at constant temperature. 
I.e. when $t_{obs}¥ \ll t_{aeff}¥$ no aging is observed and the 
dynamic response of 
the system appears stationary. In practice this applies to most 
ac-susceptibility measurements at frequencies larger than 20 Hz 
($t_{obs}¥=1/\omega \approx 10^{-2}$ s). On the other hand, 
in a dc magnetic relaxation 
experiment spanning some decades in time, the observation time of the 
experiment continuously increases and $\log t_{obs}¥$ unavoidably approaches 
$\log t_{aeff}¥$ at long enough observation times and effects of the aging 
process becomes discernible. The different times necessary to discuss 
the aging behavior are related according to: $t_{aeff}¥ \approx 
t_{w}¥+t_{obs}¥ = t_{a}¥$, where $t_{a}¥$
is the total time spent at constant temperature i.e. the age of the system.

We will continue by discussing results from ac-susceptibility experiments 
using different cooling rates and employing intermittent stops at different
temperatures during the cooling of the sample, and interpret this 
cooling behavior as well as the subsequent heating curves (always recorded 
at one and the same heating rate) in terms of an assumed domain structure 
imprinted during cooling. This discussion will be complemented by results 
from dc-magnetic relaxation experiments subsequent to cooling and heating 
procedures that closely mimic those of the ac-experiments. 
\begin{figure}
	\centerline{\epsfig{width=8.3truecm,angle=0,file=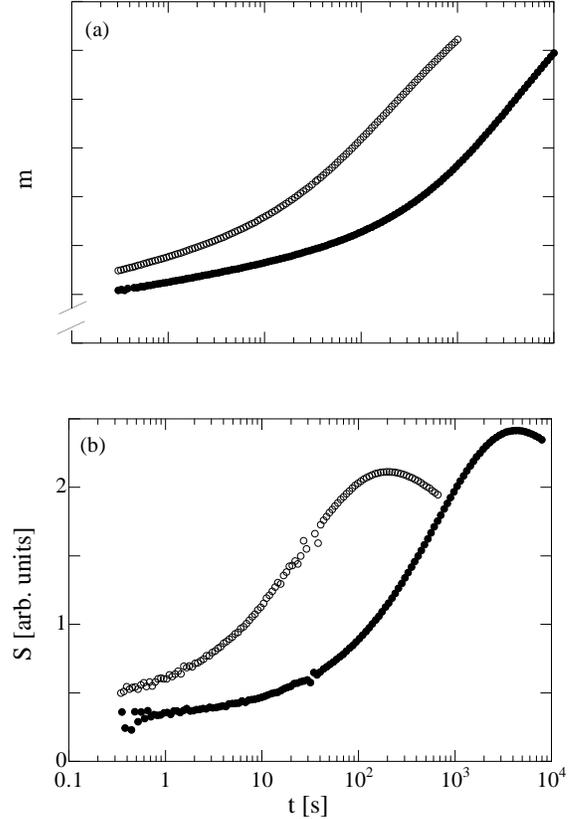}}
	\caption{Zero field cooled magnetization vs. observation time 
	at 23 K. a) $m(t)$ vs. $\log t$ b) relaxation rate $S(t)$. 
	$H_{dc}¥ = 1$ Oe, Ag(11 at\% Mn).}
\end{figure}

\subsection{ac-susceptibillity: cooling rate dependence and memory}

A simple memory effect  in the low frequency ac-susceptibility is shown 
in Fig. 4 after the following sequence has been performed. The sample is 
cooled to 23 K while measuring every 0.25 K (filled circles). At 23 K the 
sample is subjected to a wait time $t_{w}¥$ = 7 200 s during which 
$\chi''$ 
decays (cf. Fig. 3). After this wait time, the sample is further 
cooled to 20 K and then immediately  re-heated, the data on heating 
are indicated by open circles in the figure. Both the cooling and 
the heating is made with the same rate 0.25 K/min. The results from 
identical cooling and a heating procedures, but without the wait 
time at 23 K, are indicated by the full lines in Fig. 4 for 
reference. The effect of the relaxation during the wait time is 
clearly visible in the cooling curve as a dip in $\chi'' (T)$, followed 
by a rather rapid increase and approach to the reference curve 
at temperatures below 23 K on the continued cooling and, as 
expected from the droplet model, it also reappears as a 
corresponding dip in $\chi'' (T)$ centered around $T = 23$ K in the heating 
curve. The merging with the cooling reference curve on continued 
cooling can be interpreted as an effect of chaos with temperature 
\cite{BrayMoore} and a region of overlapping states, $\Delta T_{*}¥$. 
The dip on re-heating 
is a consequence of the imprinted spin configuration at $T = 23$ K on 
large length scales. I.e. when returning to this temperature all 
domain growth at lower temperatures have occurred on length 
scales $ \ll R(23$ K,7 200 s) and these are here washed out on shorter 
time scales leaving a system with an effectively equivalent domain 
configuration to the one originally obtained during the wait 
time at $T = 23$ K.

An immediate consequence of the chaos and overlap concepts is that 
the ac-susceptibility depends on the cooling rate. Larger domains have 
time to grow within the overlap region $\Delta T_{*}¥$ when 
cooling the spin-glass 
at a slower rate. Thus, the observed magnitude of $\chi'' (T)$ will be closer 
to the equilibrium level in a slow cooling process than in a fast.  
This prediction is confirmed in the inset to Fig. 4, where $\chi'' (T)$ at 
two different cooling rates, 0.25 and 0.005 K/min. are shown. The overlap 
concept is further supported by the fact that when suddenly at 25 K the 
cooling rate is increased from 0.005 to 0.25 K/min., the  cooling curve 
within a distance of $\Delta T_{*}¥ \approx$ 1 K merges with the fast
cooling curve.  
Assuming that the domain size, $R(T,t_{c}¥)$, at all temperatures in the cooling 
process is limited by a characteristic time $t_{c}¥$, which is set by the 
cooling rate and an interplay between chaos and the overlap length, 
obviously this time $t_{c}¥$ must increase with decreasing cooling rate. 
In the $\chi'' (T)$ experiments, droplet excitations on the length 
scale $L(T,1/\omega)$ 
govern the response, these excitations includes a large amount on the 
size of the domains if $1/\omega$ is of order $t_{c}¥$, but 
such excitation rapidly 
becomes rare with an increasing $t_{c}¥$. This implies as mentioned above 
that the measured susceptibility decreases with decreasing cooling rate and
that 
\begin{figure}
	\centerline{\epsfig{width=8.5truecm,angle=0,file=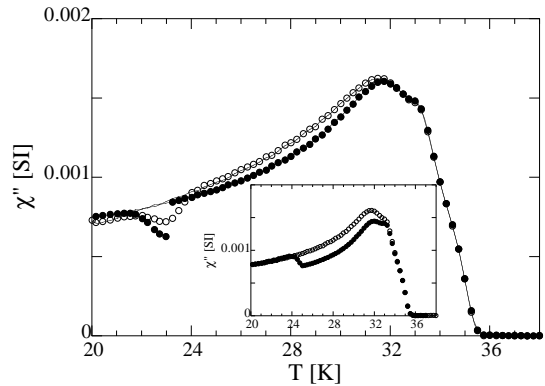}}
	\caption{$\chi''(T)$ vs. $T$ measured on cooling (solid circles) 
	and heating (open circles) at a rate of 0.25 K/min. The
	sample was intermittently kept at $T = 23$ K for $t_{w}¥ =$ 7 200 s 
	during cooling. Solid lines show the corresponding curves 
	without the interrupted cooling at 23 K. The inset shows $\chi'' 
	(T)$ 
	using two different cooling rates: 0.25 K/min. (open circles)
	and 0.005 K/min. (solid circles). The slow cooling rate is
	changed to 0.25 K/min. at 25 K. $f = 1.7$ Hz, $h_{ac}¥ = 0.1$ Oe,
	Ag(11 at\% Mn).}
\end{figure}

\noindent a change of the cooling rate must result in a crossover from 
one cooling rate characteristic curve to the other. It is of course 
also possible to cross from the fast to the slow cooling curve in a 
similar way. 

Regarding the two reference curves, measured at 0.25 K/min., in the 
main frame of Fig. 4, it may be argued that they should be almost 
identical, \cite{Nordblad} since a domain structure on continuously decreasing 
long length scales $R(T,t_{c}¥)$ is imprinted in the system on cooling 
the sample. This structure  should then also govern the 
ac-susceptibility on re-heating the sample at the same rate. 
However, as is seen in the figure, the two curves differ notably 
in magnitude. Two main factors can explain the deviation. 
Firstly, the effective heating time scale, $t_{h}¥$, is different from the 
cooling time $t_{c}¥$, although the rates are the same, 0.25 K/min. 
This difference can be understood considering the existence of an 
overlap length and the fact that the domain growth, or, equivalently, 
the domain wall movements are faster at higher temperatures. 
At any subsequent temperature, $T$, on a cooling or heating process, 
overlapping ground state configurations have already been created 
at nearby temperatures within $T \pm \Delta T_{*}¥$. Hence, the domains have 
grown to a larger size when reaching $T$ in cooling and a smaller 
size on heating, which in terms of time implies: $t_{c}¥ > t_{h}¥$. 
If the 
mapping was perfect, so that the domain structure on long length 
scales was fully intact on re-heating, the heating curve would 
still coincide with the cooling curve as long as $t_{h}¥ < t_{c}¥$, 
since we 
would probe an identical undisturbed domain structure determined
by $R(T,t_{c}¥)$ $(> R(T,t_{h}¥))$ at all temperatures. However, secondly, 
there is a reinitialisation of the gained domain structure on 
length scales that do not overlap with the created domains at a 
lower 
\begin{figure}
	\centerline{\epsfig{width=8.5truecm,angle=0,file=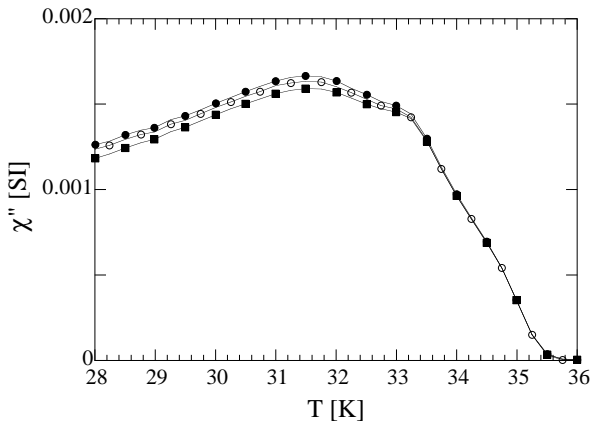}}
	\caption{$\chi'' (T)$ measured on heating at a rate of 0.25 K/min.
	The different curves are recorded after cooling the sample 
	at different cooling rates: 0.005 K/min. (solid squares), 
	0.25 K/min. (open circles) and 4 K/min. (solid circles). 
	$f = 1.7$ Hz, $h_{ac}¥ = 0.1$ Oe, Ag(11 at\% Mn).}
\end{figure}

\noindent temperature in a continuous cooling process. This implies 
that the domain structure on re-heating the sample has become 
partially reconstructed also on length scales of order $R(T,t_{c}¥)$, 
i.e. the system appears less equilibrated and the susceptibility 
attains a comparably larger magnitude. The behavior on changing 
from cooling to heating the sample requires a comment. In Fig. 4 
this change occurs at $T = 20$ K resulting in a heating reference 
curve which initially is smaller in magnitude than the cooling 
reference curve. This is explained by the fact that the sample 
spends the longest time in the temperature interval within the 
overlap region (20 K $+ \Delta T_{*}¥) \pm \Delta T_{*}¥$ first 
during cooling and then in the subsequent heating.

To elaborate further on the cooling/heating  rate dependence, 
we show in Fig. 5 three curves obtained at one and the same heating 
rate, but recorded after cooling the sample at 4 K/min., 0.25 K/min. 
and 0.005 K/min. The differences between the curves are significant 
and in accord with the discussion above that a domain structure on 
long length scales $R(T,t_{c}¥)$ is imprinted but also reinitialized 
during continuous cooling. We thus expect that the curve recorded at 
a cooling rate of 4K/min. should always yield $t_{h}¥ > t_{c}¥$ 
and  the system 
should appear least equilibrated and have the largest magnitude of 
$\chi'' (T)$. The curve obtained after cooling at 0.005 K/min. has an 
imprinted  spin structure characterized by $t_{c}¥ \gg t_{h}¥$ and  
$\chi'' (T)$ should have a substantially smaller magnitude. The curve 
measured after cooling at 0.25 K/min. is expected to
be found in-between the other two, just as is seen in the figure.

For visual clarity in the following figures, all presented ac-susceptibility 
data are subtracted by either the cooling or the heating reference curve. 
These two reference curves are obtained using a cooling/heating  rate of 
0.25 K/min. (see Fig. 4). Since the ac-susceptibility always decreases 
towards the equilibrium value when the system ages at constant temperature,
data points which are 
\begin{figure}
	\centerline{\epsfig{width=8.5truecm,angle=0,file=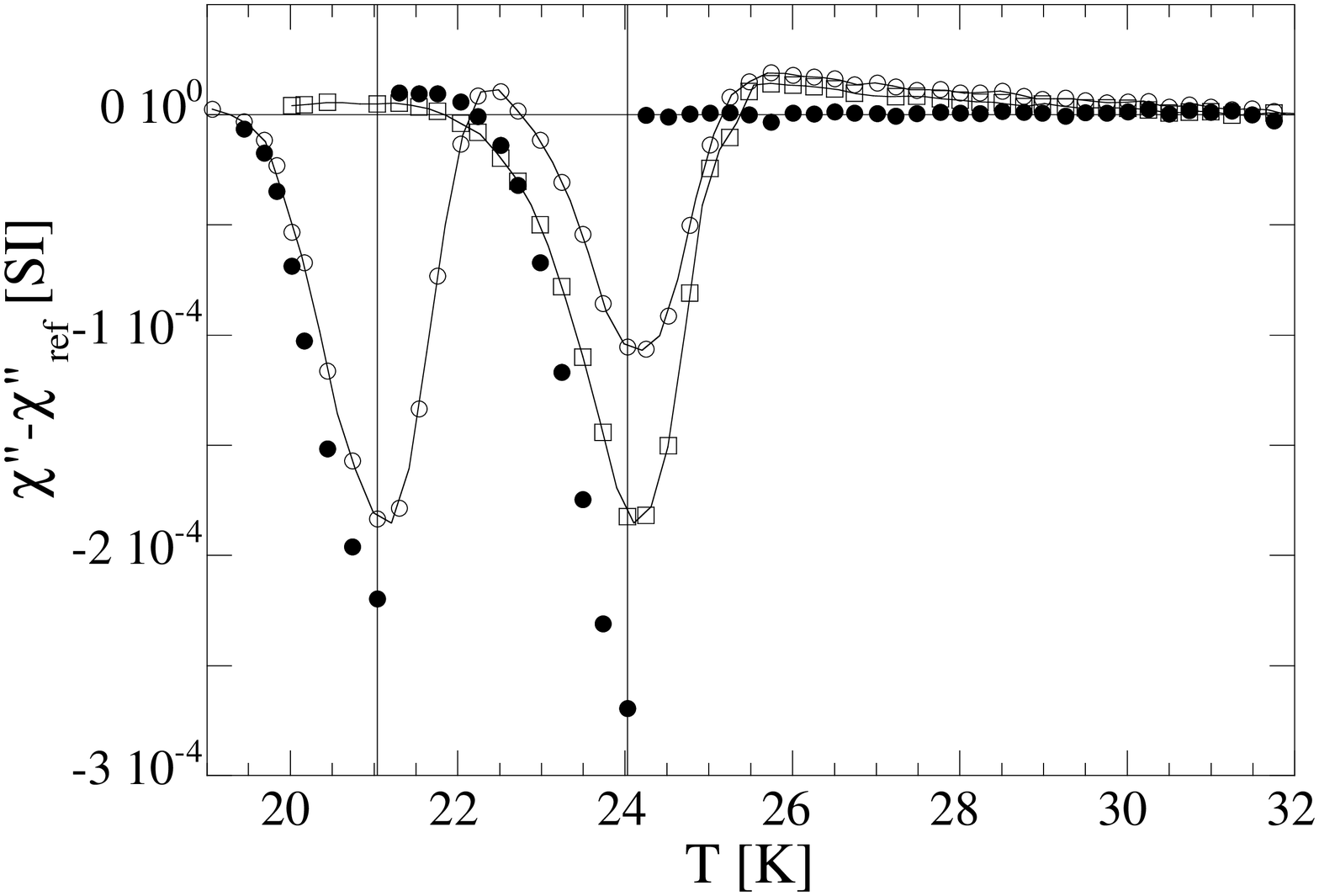}}
	\caption{$\chi''(T) - \chi''_{ref}¥(T)$ vs. $T$, measured at a cooling
	(solid symbols) or heating  rate (open symbols) of 0.25 K/min.
	The sample was kept 30 000 s at 24 and at 21 K during cooling.
	The open squares indicate an experiment where the sample only
	was kept 30 000 s at 24 K during cooling. $f = 1.7$ Hz, $h_{ac}¥ = 0.1$ Oe,
	Ag(11 at\% Mn).}
\end{figure}

\noindent positive in such plots indicate 
a state further away from equilibrium than the corresponding points 
on the reference curve. For negative values the data points are closer 
to equilibrium. Furthermore, in all these experiments, the dc-field 
is zero and the cooling is always initiated at $T = 40$ K where the 
system is paramagnetic.

As has been experimentally observed in a recent study on memory 
effects in spin-glasses \cite{Jonason} and was reproduced in Fig. 1, the 
single memory experiment presented in Fig. 4 can be extended 
to include two (or more) wait times at well separated temperatures 
during the cooling of the sample. A procedure, which in the 
subsequent heating process results in dips in the ac-susceptibility 
positioned at each of these temperatures. For our current sample, 
this intriguing behavior is illustrated in Fig. 6 where the sample 
has been subjected to two wait times of 30 000 s duration, first at 
$T = 24$ K and then at $T = 21$ K. Except at these particular temperatures 
the sample is cooled and heated with the usual rate of 0.25 K/min. As can 
be seen from the open circles in Fig. 6, the equilibration 
at each temperature gives rise to corresponding dips when re-heating the sample.

Evidently from the results discussed above, it is possible to 
imprint a spin configuration at a temperature where the spin-glass
has been allowed to age and recover this state when returning to 
the same temperature, and even to imprint and recover two or more 
equilibrated spin configurations at well separated temperatures. 
This ability excludes, as pointed out recently, \cite{Jonason} that 
new compact 
domains grow at each aging temperature. In other words, assuming 
that $|T_{1}¥ - T_{2}¥| > \Delta T_{*}¥$, it is impossible to 
imagine that compact $T_{1}¥$ domains coexist with compact $T_{2}¥$ domains.
However, by assuming that the 
equilibration process does not correspond to the growth of compact 
domains but rather, as discussed above, a removal of fractal domain 
wall structures up to the length scale $R$, the 
\begin{figure}
	\centerline{\epsfig{width=8.5truecm,angle=0,file=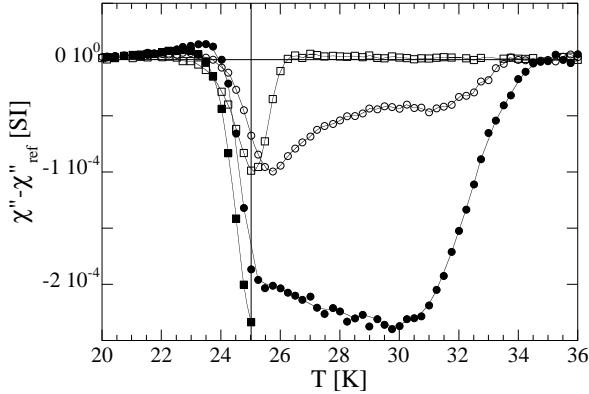}}
	\caption{$\chi''(T) - \chi''_{ref}¥(T)$ vs. $T$, measured on cooling at
	different rates (solid symbols) and heating (open symbols)
	at a rate of 0.25 K/min. Solid circles indicate a cooling
	process at 0.005 K/min. down to 25 K, where the cooling rate 
	is changed to 0.25 K/min. Open circles show the corresponding 
	curve on heating at 0.25 K/min. Solid squares below 25 K, 
	show data after having kept the sample at 25 K for 7 200 s
	using a cooling rate of 0.25 K/min. otherwise and open 
	squares the corresponding heating curve. $f = 1.7$ Hz,
	$h_{ac}¥ = 0.1$ Oe, Ag(11 at\% Mn).}
\end{figure}

\noindent results may still 
accord with the droplet model. To be more specific, the minimum 
length scale of the distance between domain walls, after the first
wait time at $T_{1}¥$ is typically $R(T_{1}¥, t_{w1}¥)$. 
Chaos with temperature \cite{BrayMoore} 
implies that this structure is random at $T_{2}¥$ on length scales larger 
than the overlap length $l(T_{1}¥ - T_{2}¥)$. After the wait time, 
$t_{w2}¥$, 
at $T_{2}¥$, the 
domain wall movements have removed fractal structures on length scales 
up to $R(T_{2}¥, t_{w2}¥)$. The spin configuration of the 
system can now be mapped 
on the equilibrium configurations at $T_{2}¥$ and at $T_{1}¥$.  
The map at $T_{2}¥$ yields 
an equilibrium system on length scales smaller than $R(T_{2}¥,t_{w2}¥)$ 
and a random 
system on larger length scales. The map at $T_{1}¥$ on the other hand shows 
random structures on length scales between $l(T_{1}¥-T_{2}¥)$ and 
$R(T_{2}¥, t_{w2}¥)$ and on 
length scales larger than $R(T_{1}¥, t_{w1}¥)$. In the real experiment, the 
configurations are less well defined and how well they are preserved in 
the thermal procedure depends on the length of the wait times, the 
cooling/heating rates and the separation between the temperatures 
$T_{1}¥$ and $T_{2}¥$. 
On re-heating the sample, the relative depth of the recovered dips in 
$\chi'' (T)$ 
in Fig. 6 mirrors how these structures have been preserved. For comparison, 
the ac-susceptibility on heating the sample from 20 K after waiting 
$t_{w}¥ =$ 30 000 s only at $T = 24$ K during the cooling, 
is included as squares in 
Fig. 6. Since this single memory dip is deeper than the corresponding 
double memory dip it can be concluded that during the wait time at 
$T = 21$ K 
the first imprinted  spin configuration at the preceding aging at 
$T = 24$ K 
has been partly over written, i.e. the system has been partly reinitialized. 
It is also worth to notice that the relative depth of the dip on re-heating 
the sample is temperature dependent; with increasing temperature, in otherwise 
identical single memory experiments, the dip becomes more and more 
shallow compared to the depth obtained during the original wait 
time at $T_{m}¥$ on cooling (see also Fig. 7).

A partial reinitialisation, is also evidenced from the positive 
values of the relative ac-susceptibility plotted in Fig. 6 a few 
Kelvin above and below the dip at 24 K, signaling an apparently 
less equilibrated system as compared to the reference. Before 
discussing the origin of this behavior more thoroughly, it is 
advantageous to return to the very slow cooling experiment. 
In Fig. 7 the ac-susceptibility during cooling with 0.005 K/min., 
subtracted by the corresponding data obtained with the cooling 
rate 0.25 K/min., is represented by filled circles. When passing 
the temperature $T = 25$ K the cooling rate is changed to 0.25 K/min. 
resulting in a quick increase of the susceptibility towards the 
reference level. In fact it even crosses the reference curve during 
the proceeding cooling to $T = 20$ K. The ac-susceptibility of the 
subsequent heating, at rate 0.25 K/min., are shown by the open 
circles. As can be expected when considering the overlap length 
and a decrease of the reinitialisation just when the cooling rate 
is increased, the system looks most equilibrated in a temperature 
interval $T=(25$ K$+ \Delta T_{*}¥) \pm \Delta T_{*}¥$. 
At higher temperatures, the spin 
configuration appears less equilibrated than the original slow cooling 
curve, but still closer to equilibrium than the reference heating 
curve (which is measured at the same heating rate, 0.25 K/min.). 
This behavior mirrors both the continuously imprinted equilibrated 
spin configurations obtained during slow cooling and the simultaneous 
partial reinitialisation  of these configurations which occurs at 
lower temperatures. It must hence also be concluded that the 
equilibrated length scales $R(T,t_{c}¥)$ obtained at each temperature 
in the slow cooling process are partly preserved during the whole 
experiment. For comparison we have in Fig. 7 included a single
memory curve recorded at a cooling/heating rate of 0.25 K/min. 
and waiting 7 200 s at 25 K during cooling. Filled squares show 
the cooling curve and open squares the heating curve. It may be 
noticed that this dip is centered around $T=25$ K as expected.

Fig. 6 and 7 show that a long wait time at constant temperature also 
affects the ac-response at surrounding temperatures, so that the 
spin-glass appears less equilibrated than in the reference 
procedure at temperatures separated more than $\Delta T_{*}¥$ from the 
temperature where the aging took place. As discussed above, in the
continuous cooling case the typical equilibrated domain 
size $R(T,t_{c}¥)$ is governed by the effective cooling time $t_{c}¥$, which 
also sets an effective age and thus the 
magnitude of $\chi'' (T,t_{c}¥)$. When 
the cooling is stopped at $T_{m}¥$, the domains can grow unrestricted, 
and after the wait time they have reached a size $R(T_{m}¥,t_{w}¥)$. When 
cooling is recaptured, this new domain size is adequate as long as 
the overlap length, $l(T_{m}¥-T) > R(T,t_{w}¥)$, however, at temperatures just 
outside $T \le (T_{m}¥-\Delta T_{*}¥)$ the typical
domain size is determined by the overlap length 
\begin{figure}
	\centerline{\epsfig{width=8.5truecm,angle=0,file=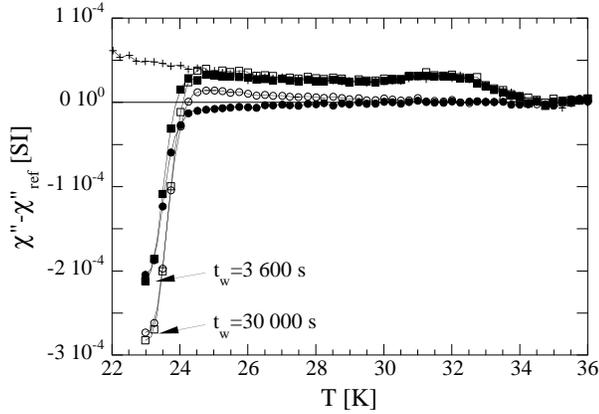}}
	\caption{$\chi''(T) - \chi''_{ref}¥(T)$ vs. $T$, measured on 
	heating at 0.25 K/min., after the sample has been cooled 
	to 23 K at 0.25 K/min. (circles) or 4 K/min. (squares) and 
	kept at 23 K for 3 600 s (solid symbols) or 30 000 s 
	(open symbols). The pluses show a heating curve measured after 
	cooling the sample at 4 K/min. without a stop at 23 K.
	$f = 1.7$ Hz, $h_{ac}¥=0.1$ Oe, Ag(11 at\% Mn).}
\end{figure}

\noindent $l(T_{m}¥-T) \ll R(T,t_{w}¥)$. (The overlap length, $l(\Delta T)$, is a 
very rapidly decreasing function with $\Delta T$).  At lower temperatures,  
the cooling time again becomes the governing parameter. However, 
just outside the overlap region, $T_{m}¥-\Delta T_{*}¥$,
the actual cooling time, $t_{ac}¥$, 
i.e. the time the sample has been kept within a region where the 
growing domain size overlaps with the domain size achieved at temperatures 
just above, is of course considerably shorter than the cooling time, 
$t_{c}¥$, of the continuous reference cooling procedure and thus  
$R(T,t_{ac}¥) < R(T,t_{c}¥)$. During the continued cooling, $t_{ac}¥$
increases, but a temperature decrease at least of order 
$\Delta T_{*}¥$ is needed to regain $t_{ac}¥ \approx t_{c}¥$ 
and a typical domain size equivalent to that of 
the continuous cooling procedure. These arguments suggest, in 
accord with the experiments, that the disturbed $\chi'' (T)$-curve at 
temperatures below $T_{m}¥$ first should increase towards and even cross 
the reference level as the temperature is decreased beyond $\Delta 
T_{*}¥$, 
go through a maximum deviation and then slowly approach $\chi''_{ref}¥$ further
away from $T_{m}¥$. Similar arguments as discussed above can also be 
applied to the heating curve at temperatures above $T_{m}¥$, i.e. the dip 
should be succeeded by a temperature region outside $T_{m}¥ + \Delta 
T_{*}¥$ where the 
relative susceptibility should become positive. The longer time the 
sample is kept at $T_{m}¥$ the deeper the dip and the larger the 
positive levels outside the overlap region will become. This conclusion is 
further elucidated in an experiment where the sample is cooled, either
with the rate 0.25 K/min. or 4 K/min., to the temperature 23 K where 
it is subjected to different wait times before it is heated with the 
rate 0.25 K/min. In Fig. 8 the ac-susceptibilities, measured on  
 heating, are indicated by filled symbols after the wait time, 
$t_{w}¥ =$ 3 600 s. Open symbols indicate the corresponding curves after 
a wait time, $t_{w}¥ =$ 30 000 s. To separate the two cooling rates, 
squares are used to 
\begin{figure}
	\centerline{\epsfig{width=8.5truecm,angle=0,file=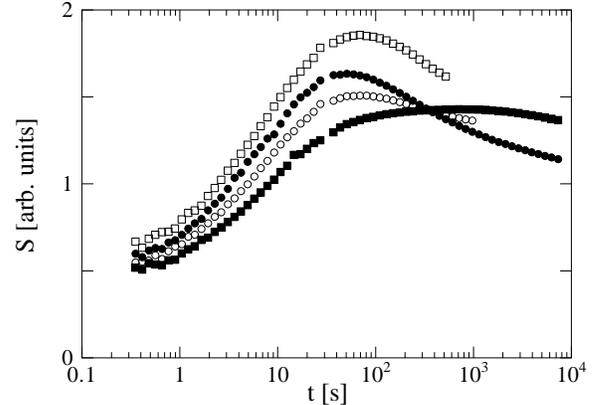}}
	\caption{Relaxation rate $S$ vs. $\log t$, measured at 25 K after
	having cooled the sample at 0.25 K/min. to 23 K and keeping it
	there a wait time tw before re-heating it to 25 K and applying 
	the dc-field after 10 s at constant temperature. 
	$t_{w}¥ = 0$ (filled squares), 3 600 s (open circles) and 30 000 s 
	(solid circles). A reference curve measured after cooling 
	the sample at 4 K/min. immediately to 25 K, and applying the
	field after 10 s is indicated by open squares.
	$H_{dc}¥ = 1$ Oe, Ag(11 at\% Mn).}
\end{figure}

\noindent symbolize the faster one while circles are used 
for the slower. Also included in Fig. 8 is a reference curve 
obtained during heating after cooling to 20 K with the fast rate, 
4 K/min. As can be seen, the system appears more reinitialized, 
at high enough temperatures, after a waiting time of 30 000 s. 
The effect is most pronounced in the case where a slow cooling 
rate was used. A wait time of 3 600 s is apparently too short 
realize a measurable reinitialisation.

\subsection{Relaxation rate experiments}
The usefulness of this fractal domain picture can be further 
examined by means of relaxation measurements where the observation 
time spans several decades. A zero-field-cooled relaxation 
experiment  can be designed to give complementary information on 
the reinitialisation observed in the ac-susceptibility curves 
of Fig. 7. In these measurements, as in the ac-experiments, the
sample is cooled with a rate of 0.25 K/min. to 23 K where it is 
subjected to a wait time, $t_{w}¥$. The sample is then heated at the 
rate 0.25 K/min. but only to 25 K where the  sample is kept at
constant temperature and after a time of 10 s a magnetic field of
1 Oe is applied. The subsequent relaxation rate is shown in 
Fig. 9 for the three different wait times, $t_{w}¥ = 0$ 
(filled squares), 3 600 s (open circles), and 30 000 s (filled circles). 
For reference purpose, the corresponding rate obtained after 
cooling the sample at 4 K/min. directly to 25 K and waiting 
there 10 s before the field is applied, is included and 
indicated by open squares. It should also be mentioned that 
a similar
\begin{figure}
	\centerline{\epsfig{width=8.5truecm,angle=0,file=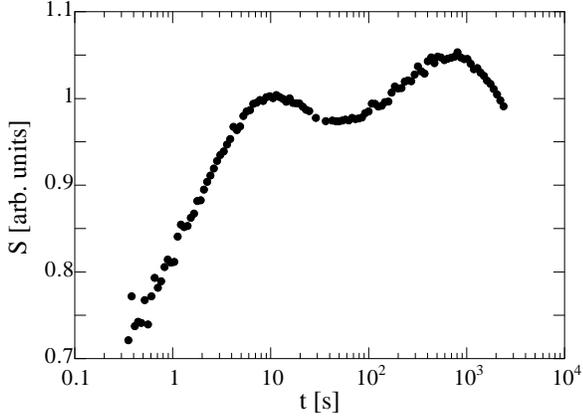}}
	\caption{Relaxation rate $S$ vs. $\log t$, measured at 31 K. 
	The sample was cooled to 31 K, kept there a wait 
	time 1 000 s, then cooled to 29.7 K, kept there 3 000 s and
	finally re-heated to 31 K, where the dc-field was applied. 
	$H_{dc}¥ = 1$ Oe, Ag(11at\% Mn).}
\end{figure}

\noindent  measurement  where the sample is directly cooled 
to 25 K using a cooling rate of 0.25 K/min. gives a relaxation
rate curve that is closely equal to the curve where a zero second
wait time at 23 K is used (filled squares). The prime aging 
characteristic of the magnetic relaxation in spin-glasses is
an inflection point in the relaxation rate at an observation 
time closely equal to the wait time before the magnetic field 
is applied, $t_{w}¥$. A rather slow cooling rate, like 0.25 K/min., 
yields a broad and ill defined maximum in the relaxation rate 
at an observation time that may be called effective wait time or
age. The three curves taken after a cooling rate of 0.25 K/min. 
in Fig. 9 show a continuous development towards the rapidly cooled
(4K/min.) reference curve with increasing wait time at 23 K. 
In terms of the droplet model and in agreement with the discussions
above, the behavior may be interpreted as follows; a longer wait
time at $T = 23$ K should create equilibrated regions on larger and 
larger length scales, length scales that should approach the typical
domain size achieved during cooling at higher temperatures. At 25 K,
which on our experimental time scales is out of the overlap region,
these regions looks random, i.e. with increasing wait time at 23 K, 
the system should appear more and more reinitialized at $T = 25$ K, 
just as Fig. 9 shows.

The established way to show the possibility for a spin-glass to 
simultaneously carry several characteristic length scale, at one 
temperature, is by temperature cycling experiments. \cite{Granberg} 
The relaxation 
rate result presented in Fig. 10 is obtained by first cooling the sample 
to $T_{1}¥ = 31$ K where it is subjected to a wait time of 1 000 s. The
temperature is thereafter changed to $T_{2}¥ = 29.7$ K where it is kept 
for $t_{w2}¥ =$ 3 000 s. When returning to $T_{1}¥$ again after this temperature
cycling, diluted regions of size $R(T_{2}¥, t_{w2}¥)$ have been reinitialized
due to the domain wall movements at $T_{2}¥$. The effect of a third wait
time $t_{w3}¥ = 10$ s at $T_{1}¥$ is a system which initially equilibrates 
\begin{figure}
	\centerline{\epsfig{width=8.5truecm,angle=0,file=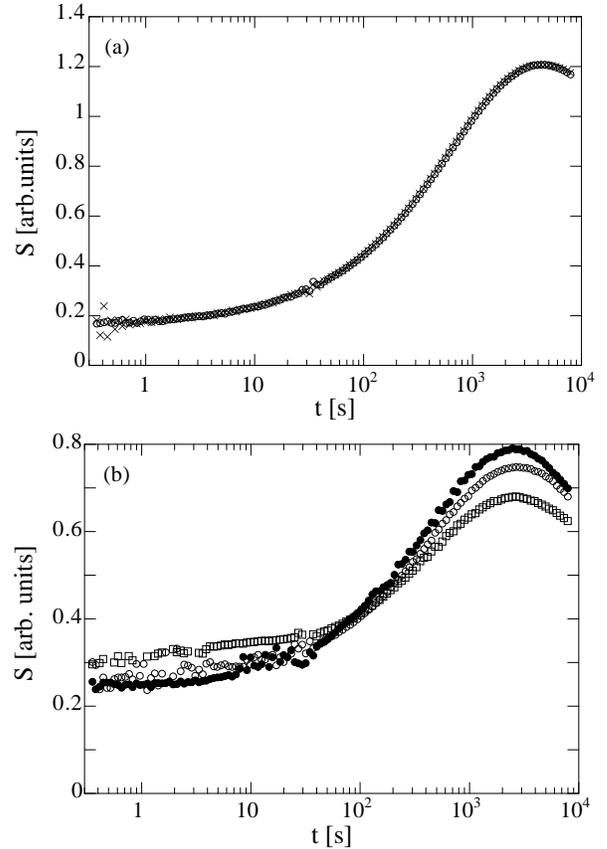}}
	\caption{a) Relaxation rate $S$ vs. $\log t$, measured at 27 K after
	a wait time of 3 000 s. Crosses, the sample was immediately 
	cooled to 27 K, open circles, the sample was intermittently kept 
	at 31 K for 3 000 s during cooling. b) Relaxation rate $S$ vs. 
	$\log t$, measured at 31 K after having heated the sample back 
	from 27 K and immediately applying the dc-field (open squares).
	Open circles shows a corresponding curve where the sample was
	kept at 23 K for 3 000 s instead of at 27 K. Solid circles shows
	the relaxation rate after cooling the sample directly to 31 K 
	and wait 3 000 s before applying the field. $H_{dc} = 1$ Oe,
	Ag(11at\% Mn).}
\end{figure}

\noindent in these
reinitialized regions up to length scales $R(T_{1}¥, t_{w3}¥)$. By choosing
the temperature step and the wait times carefully it is possible
to fulfill the condition
$R(T_{1}¥,t_{w3}¥) < R(T_{2}¥,t_{w2}¥) < R(T_{1}¥,t_{w1}¥)$. The
two maxima in the relaxation rate in Fig. 10 then correspond to
the non-equilibrium response due to the remaining two relevant 
characteristic length scales at $T_{1}¥$, $R(T_{1}¥, t_{w3}¥)$ 
and $R(T_{1}¥, t_{w1}¥)$.

It is also possible to show that several length scales can coexist in 
the spin-glass at different temperatures with relaxation measurements. 
Indicated by crosses in Fig. 11a is the relaxation rate for the
temperature 27 K after a wait time of 3 000 s. Also shown in the same 
figure, indicated by circles, is the relaxation rate after the 
sample has spent 3 000 s at 31 K before the aging at 27 K. 
No difference in the measured relaxation rate can be resolved
between these two procedures. A length scale, caused by the wait 
time at 31 K, is however hidden in the second case. In order to 
reveal this, the sample is heated back to 31 K and immediately probed 
in a magnetic field. The subsequent relaxation rate is indicated by 
squares in Fig 11b. Also shown by open circles in this figure, is
the corresponding relaxation rate but with the second aging at 23 K 
instead of 27 K. As a reference, the relaxation rate probed directly 
after the initial wait time at 31 K is shown as filled circles. 
Basically, all three curves show an age of the system equal to 
3 000 s only differing in the magnitude of the rate. This experiment, 
similar to the ac double memory experiment, shows that a spin-glass
always recalls what has happened during cooling when it is re-heated 
to (or through) the temperature of an interrupted cooling process.

\section{Conclusions}
The non-equilibrium behavior of the dynamic susceptibility of a 
3d spin-glass has been measured after specific thermal procedures
which have been designed to expose consequences of a corresponding 
domain growth as predicted  in the Fisher and Huse droplet model 
of spin-glass dynamics. \cite{FisherHuse} 
A good qualitative agreement between 
these experimental results and the model predictions is found, 
which give support for the relevance of this real space model. 
However, there remain important questions to be answered. One 
crucial point within the model is to obtain an understanding of 
the difference between the initial fractal domain configuration after
a quench and a corresponding snap shot of the equilibrated system
with droplet excitations on all length scales. Is it a difference of
the fractal dimension of the instantaneous ÔdomainÕ walls in the 
two cases that signifies the difference? A prime remaining issue is 
also to theoretically establish if the concept of fractal domains on 
many length scales can be valid for a model Ising as well as for a
real 3d spin-glass.

\acknowledgements
Financial support from the Swedish Natural Science Research 
Council (NFR) is acknowledged. Yvonne Andersson and Hui-Ping
Liu are acknowledged for invaluable assistance in preparing 
our AgMn sample.

\end{multicols}

\end{document}